\documentclass[twocolumn,showpacs,preprintnumbers,amsmath,amssymb]{revtex4}

\usepackage{graphicx}% Include figure files
\usepackage{bm}% bold math

\begin{document}

%\preprint{for submission}

\title{Microwave Surface-Impedance Measurements of the Magnetic Penetration Depth in Single Crystal Ba$_{1-x}$K$_x$Fe$_2$As$_2$ Superconductors: Evidence for a Disorder-Dependent Superfluid Density}

\author{K. Hashimoto,$^{1}$ T. Shibauchi,$^{1}$ S. Kasahara,$^{2,3}$ K. Ikada,$^{1}$ S. Tonegawa,$^{1}$ T. Kato,$^{1}$ R. Okazaki,$^{1}$ \\
C.~J. van der Beek,$^{4}$ M. Konczykowski,$^{4}$ 
H. Takeya,$^{3}$ K. Hirata,$^{3}$ T. Terashima,$^{2}$ and Y. Matsuda$^{1,4}$}

\affiliation{$^1$Department of Physics, Kyoto University, Sakyo-ku, Kyoto 606-8502, Japan\\
$^2$Research Center for Low Temperature and Materials Sciences, Kyoto University, Sakyo-ku, Kyoto 606-8502, Japan\\
$^3$National Institute for Materials Science, 1-2-1 Sengen, Tsukuba, Ibaraki 305-0047, Japan\\
$^4$Laboratoire des Solides Irradi\'es, Ecole Polytechnique, 91128 Palaiseau cedex, France
}

\date{\today}

\begin{abstract}
We report high-sensitivity microwave measurements of the in-plane penetration depth $\lambda_{ab}$ and quasiparticle scattering rate $1/\tau$ in several single crystals of hole-doped Fe-based superconductor Ba$_{1-x}$K$_x$Fe$_2$As$_2$ ($x\approx 0.55$). While power-law temperature dependence of $\lambda_{ab}$ with the power $\sim 2$ is found in crystals with large $1/\tau$, we observe exponential temperature dependence of superfluid density consistent with the existence of fully opened two gaps in the cleanest crystal we studied. The difference may be a consequence of different level of disorder inherent in the crystals. We also find a linear relation between the low-temperature scattering rate and the density of quasiparticles, which shows a clear contrast to the case of $d$-wave cuprate superconductors with nodes in the gap. These results demonstrate intrinsically nodeless order parameters in the Fe-arsenides. 
\end{abstract}

%74.25.Nf 	Response to electromagnetic fields (nuclear magnetic resonance, surface impedance, etc.)
%74.70.-b 	Superconducting materials (for cuprates, see 74.72.-h)
%74.20.Rp 	Pairing symmetries (other than s-wave)
%74.25.Fy 	Transport properties (electric and thermal conductivity, thermoelectric effects, etc.)

\maketitle

%\section{Introduction}

The discovery of high-$T_c$ superconductivity in Fe-pnictides \cite{Kam08} has attracted tremendous interests both experimentally and theoretically. The `mother' materials have antiferromagnetic spin-density-wave order \cite{Cru08} and the superconductivity appears by doping charge carriers, either electrons or holes. Such carrier doping induced superconductivity resembles high-$T_c$ cuprates, but one of the most significant differences is the multiband electronic structure having electron and hole pockets in the Fe-based superconductors. Unconventional superconducting pairings, most notably the sign-reversing $s_\pm$ state, have been suggested by several theories \cite{Maz08} featuring the importance of the nesting between the hole and electron bands. This is also in sharp contrast to other multiband superconductors such as MgB$_2$, where the coupling between the different bands is very weak. Thus the most crucial is to clarify the novel multiband nature of superconductivity in this new class of materials. 

In this context, identifying the detailed structure of superconducting order parameter, particularly the presence or absence of nodes in the gap function, is of primary importance. In the electron-doped $Ln$FeAs(O,F) or `1111' (where $Ln$ is Lanthanide ions), while several experiments using polycrystals \cite{Nak08,Sha08} suggest nodes in the gap, the tunnelling measurements \cite{Che08}, the magnetic penetration depth \cite{Has08}, and angle resolved photoemission (ARPES) \cite{Kon08} consistently indicate fully gapped superconductivity. In the hole-doped Ba$_{1-x}$K$_x$Fe$_2$As$_2$ or `122' \cite{Rot08}, however, experimental situation is controversial: ARPES \cite{Zha08,Din08,Evt08} and lower critical field measurements \cite{Ren08} support nodeless gaps while $\mu$SR measurements \cite{Gok08} suggest the presence of line nodes in the gap function. Also, recent penetration depth measurements suggest superconductivity with point nodes in Ba(Fe$_{0.93}$Co$_{0.07})_2$As$_2$ \cite{Gor08}. 

Such controversies may partly come from the different samples with various degrees of disorder in these reports. Indeed, in the unconventional $s_\pm$ state, the quasiparticle excitation spectrum is found to be sensitive to the pair breaking and interband scattering \cite{Par08,Sen08}. Also, inhomogeneity or weak-links may affect the superfluid density \cite{Hal92}, which is a direct measure of low-energy quasiparticle excitations. So experiments in single crystals with well-characterized quality are needed to elucidate how disorder affects the quasiparticle excitations and what is the intrinsic structure of the gap in the clean limit. 

Here we present surface impedance measurements in several single crystals of Ba$_{1-x}$K$_x$Fe$_2$As$_2$ ($x\approx 0.55$). In this material, disorder may be caused by microscopic inhomogeneous content of K, which is reactive with moisture or oxygen. The degree of disorder can be quantified by the quasiparticle scattering rate $1/\tau$. We find that the temperature dependence of the in-plane penetration depth $\lambda_{ab}(T)$ is sample dependent, which can account for some of the controversies in the previous reports: crystals with large $1/\tau$ tend to exhibit power-law temperature dependence of $\lambda_{ab}(T)$. In the best crystal with the smallest $1/\tau$, we obtain strong evidence for the nodeless superconductivity having at least two different gaps. 

%\section{Experimental}
Single crystals of Ba$_{1-x}$K$_x$Fe$_2$As$_2$ were grown by a self-flux method \cite{Kas08}. High purity starting materials were heated up to 1190$^\circ$C under Ar atmosphere, and then cooled down at a rate of 4$^\circ$C/hours, followed by a quench at 850$^\circ$C. Energy dispersive X-ray (EDX) analysis reveals the doping level $x=0.55(2)$ \cite{EDX}, which is consistent with the $c$-axis lattice constant $c=1.341(3)$~nm determined by X-ray diffraction \cite{Luo08}. Bulk superconductivity is characterized by the magnetization measurements using a commercial magnetometer. In-plane microwave surface impedance $Z_s=R_s+{\rm i}X_s$ %, where $R_s$ ($X_s$) is the surface resistance (reactance), 
is measured in the Meissner state by using a sensitive superconducting cavity resonator \cite{Shi94,Shi07,Has08}. In our frequency range $\omega/2\pi\approx 28$~GHz, the conductivity $\sigma=\sigma_1-{\rm i}\sigma_2$ can be extracted from $Z_s(T)$ through the relation valid for the skin-depth regime: 
\begin{equation}
Z_s=R_s+{\rm i}X_s=\left(\frac{{\rm i}\mu_0\omega}{\sigma_1-{\rm i}\sigma_2}\right)^{1/2}.
\label{impedance}
\end{equation}
In the superconducting state, the surface reactance is a direct measure of the superfluid density $n_s$ via $X_s(T)=\mu_0\omega\lambda_{ab}(T)$ and $\lambda_{ab}^{-2}(T)=\mu_0n_s(T)e^2/m^*$. In the normal state, $\sigma_1=ne^2\tau/m^*\gg\sigma_2$ gives $R_s(T)=X_s(T)=(\mu_0\omega/2\sigma_1)^{1/2}$ from Eq.~(\ref{impedance}), where $n$ is the total density of carriers with effective mass $m^*$. This equality can be used to determine $X_s(0)/X_s(T_c)$, which allows us to determine $\lambda_{ab}(T)/\lambda_{ab}(0)$ without any assumptions \cite{Shi94}. This also gives us estimates of the normal-state scattering rate $1/\tau=1/\mu_0\sigma_1\lambda_{ab}^2(0)=2\omega(X_s(T)/X_s(0))^2$, which quantifies the degrees of disorder for each sample [see Table~\ref{table1}]. Below $T_c$, the real part of conductivity $\sigma_1$ is determined by the quasiparticle dynamics, and in the simple two-fluid model, which is known to be useful in cuprate superconductors \cite{Bon07,Bon94}, $\sigma_1$ is related to the quasiparticle scattering time $\tau$ through $\sigma_1=(n-n_s)e^2\tau/m^*(1+\omega^2\tau^2)$. 

%\section{Results and Discussion}

\begin{table}[b]
\caption{\label{table1} Properties in the crystals we studied. In this study, the transition temperature $T_c$ is evaluated from the extrapolation of superfluid density $n_s\rightarrow0$.}
\begin{ruledtabular}
\begin{tabular}{ccccccc}
sample &size ($\mu$m$^3$)&$T_c$ (K)& $\frac{1}{\tau(40~{\rm K})}$ (s$^{-1}$)& $\frac{1}{\tau(15~{\rm K})}$ (s$^{-1}$)\\
\hline
\#1& $320\times 500\times 100$ & 26.4(3) & $27(3)\times10^{12}$ & $1.2(1)\times10^{12}$ \\
\#2& $300\times 500\times 80$ & 25.0(4) & $21(2)\times10^{12}$ & $1.0(1)\times10^{12}$ \\
\#3& $100\times 180\times 20$ & 32.7(2) & $7.8(5)\times10^{12}$ & $0.5(1)\times10^{12}$ 
\end{tabular}
\end{ruledtabular}
\end{table}

%%%%%%%%%%%%%%%%%%%%%%%%%%%%%%%%%%%FIG 1%%%%%%%%%%%%%%%%%%%%
\begin{figure}[t]
\includegraphics[width=80mm]{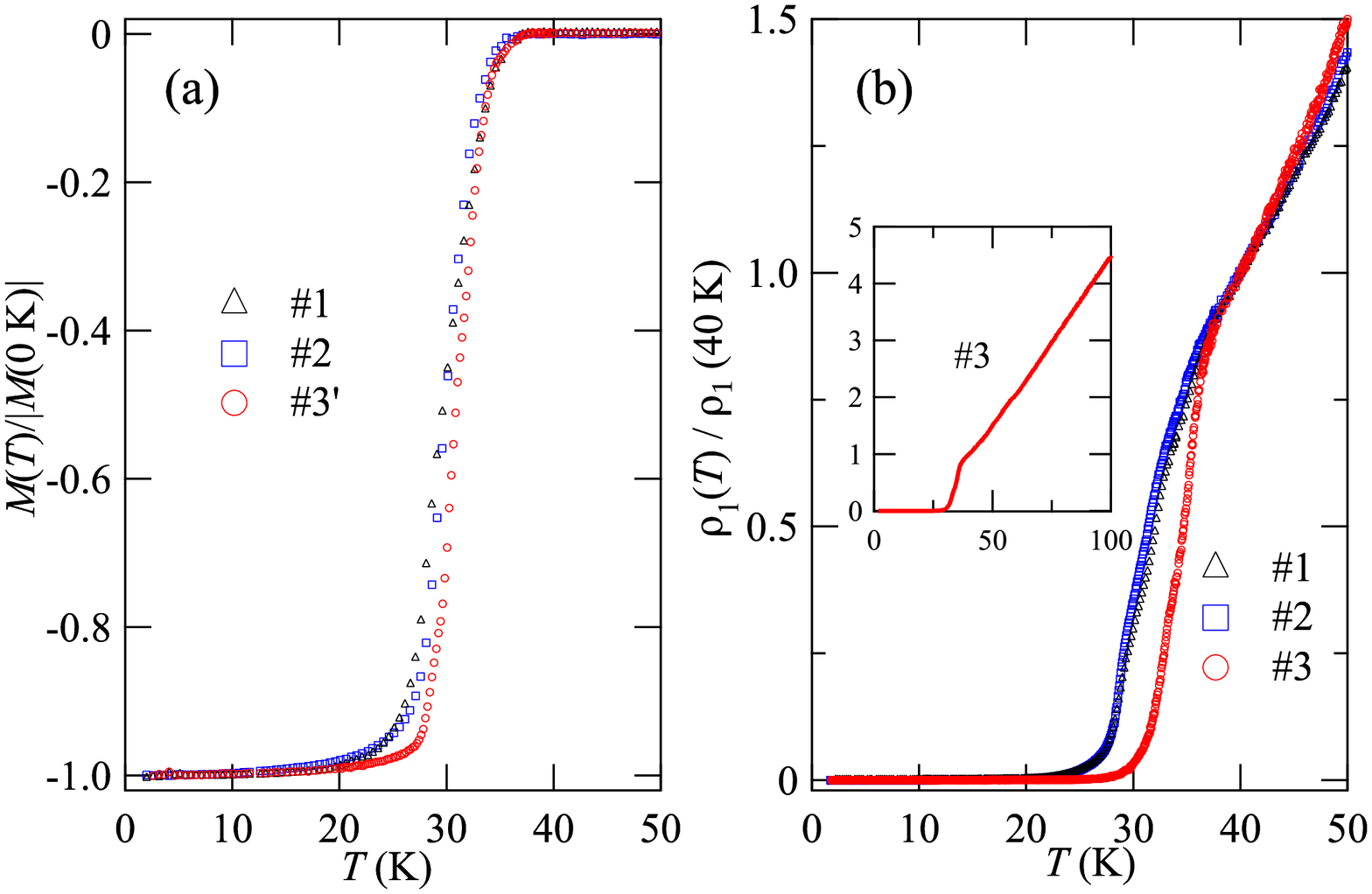}%
\caption{(color online). (a) Temperature dependence of dc magnetization in the zero-field-cooling condition under a 20-Oe field along the $c$ axis. (b) Temperature dependence of the normalized 28-GHz microwave resistivity $\rho_1(T)/\rho_1(40$~K). Inset: the same plot for crystal \#3 up to 100 K.}
\label{MT}
\end{figure}
%%%%%%%%%%%%%%%%%%%%%%%%%%%%%%%%%%%FIG 1%%%%%%%%%%%%%%%%%%%%

Figure~\ref{MT}(a) shows the temperature dependence of magnetization $M$. In the samples studied here, the low-temperature $M$ (below $\sim10$~K) is independent of temperature, which ensures the bulk nature of superconductivity. However, the superconducting transition width varies slightly from sample to sample. Since this is likely related to the microscopic inhomogeneity of K content near the surface, which can be enhanced upon exposure to the air, we carefully cleave both sides of the surface of crystal \#2 and cut into smaller size (crystal \#3). For \#3, the microwave measurements are done with minimal air exposure time, and another piece of cleaved crystal (\#3') is used to measure $M(T)$. In Fig.~\ref{MT}(b) we compare the temperature dependence of normal-state microwave resistivity $\rho_1=1/\sigma_1={2R_s^2/\mu_0\omega}$ [see Eq.~(\ref{impedance})] for 3 samples. Note that such microwave measurements provides one of the most severe quality checks for superconductors \cite{Hal92}; additional quasiparticle excitations by the applied 28-GHz ($\sim1.3$~K) microwave should intrinsically broaden the transition width compared with the dc resistivity. It is now clear that the cleavage dramatically improves the sample quality, and crystal \#3 exhibits the sharpest transition and the lowest $1/\tau$ [see Table~\ref{table1}]. The ratio $\rho_1(100$~K$)/\rho_1(T_c)$ exceeds 4 [inset of Fig.~\ref{MT}(b)], which is larger than that of less doped crystals with $x\lesssim0.4$ \cite{Luo08}. 

%%%%%%%%%%%%%%%%%%%%%%%%%%%%%%%%%%%FIG 2%%%%%%%%%%%%%%%%%%%%
\begin{figure}[t]
\includegraphics[width=85mm]{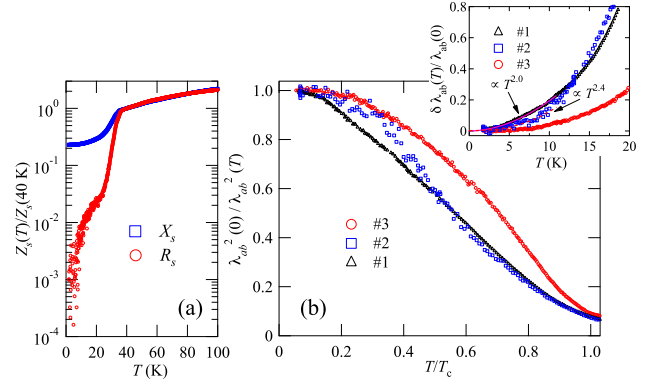}%
\caption{(color online). (a) Temperature dependence of the surface resistance $R_s$ and $X_s$ in crystal \#3. (b) Normalized superfluid density $\lambda_{ab}^2(0)/\lambda_{ab}^2(T)$ for 3 samples with different normal-state scattering rates (see Table~\ref{table1}). Inset shows the low-temperature change in the penetration depth.}
\label{Z_s}
\end{figure}
%%%%%%%%%%%%%%%%%%%%%%%%%%%%%%%%%%%FIG 2%%%%%%%%%%%%%%%%%%%%

In Fig.~\ref{Z_s}(a), we show typical temperature dependence of $Z_s(T)/Z_s$(40~K). As expected from Eq.~(\ref{impedance}), the normal-state temperature dependencies of real and imaginary parts are identical. The strong temperature dependence above $T_c$ allows us to determine precisely the offset of $X_s(0)/X_s$(40~K), from which we are able to obtain $n_s(T)/n_s(0)=\lambda_{ab}^2(0)/\lambda_{ab}^2(T)$ as demonstrated in Fig.~\ref{Z_s}(b). We find that crystals with large scattering rates exhibit strong temperature dependence of superfluid density at low temperatures, which mimics the power-law temperature dependence of $n_s(T)$ in $d$-wave superconductors with nodes. As shown in the inset of Fig.~\ref{Z_s}(b), the low-temperature change in the penetration depth $\delta\lambda_{ab}(T)=\lambda_{ab}(T)-\lambda_{ab}(0)$ can be fitted to $T^2$ and $T^{2.4}$ dependence in crystal \#1 and \#2 respectively. However, the data in cleaner samples show clear flattening at low temperatures. This systematic change suggests that the superfluid density is quite sensitive to disorder in this system and disorder promotes quasiparticle excitations significantly. It is tempting to associate the observed effect with unconventional superconductivity with sign reversal such as the $s_\pm$ state \cite{Maz08}, where impurity scattering may induce in-gap states in clear contrast to the conventional $s$-wave superconductivity \cite{Par08,Sen08}. Indeed, $T_c$ determined by $n_s \to 0$ is noticeably reduced for samples with large $1/\tau$ [Table \ref{table1}], consistent with theoretical studies \cite{Sen08}. Also, the impurity-induced change of $\delta\lambda_{ab}(T)$ from the exponential to $T^2$ dependence has been suggested theoretically \cite{Vor09}, in good correspondence with our results. At present stage, however, the microscopic nature of disorder inherent in our crystals is unclear, and a more controlled way of varying degrees of disorder is needed for further quantitative understanding of the impurity effects in Fe-based superconductors. In any case, our results can account for some of the discrepancies in the reports of superfluid density in Fe-arsenides \cite{Gok08}. 

%%%%%%%%%%%%%%%%%%%%%%%%%%%%%%%%%%%FIG 3%%%%%%%%%%%%%%%%%%%%
\begin{figure}[t]
\includegraphics[width=85mm]{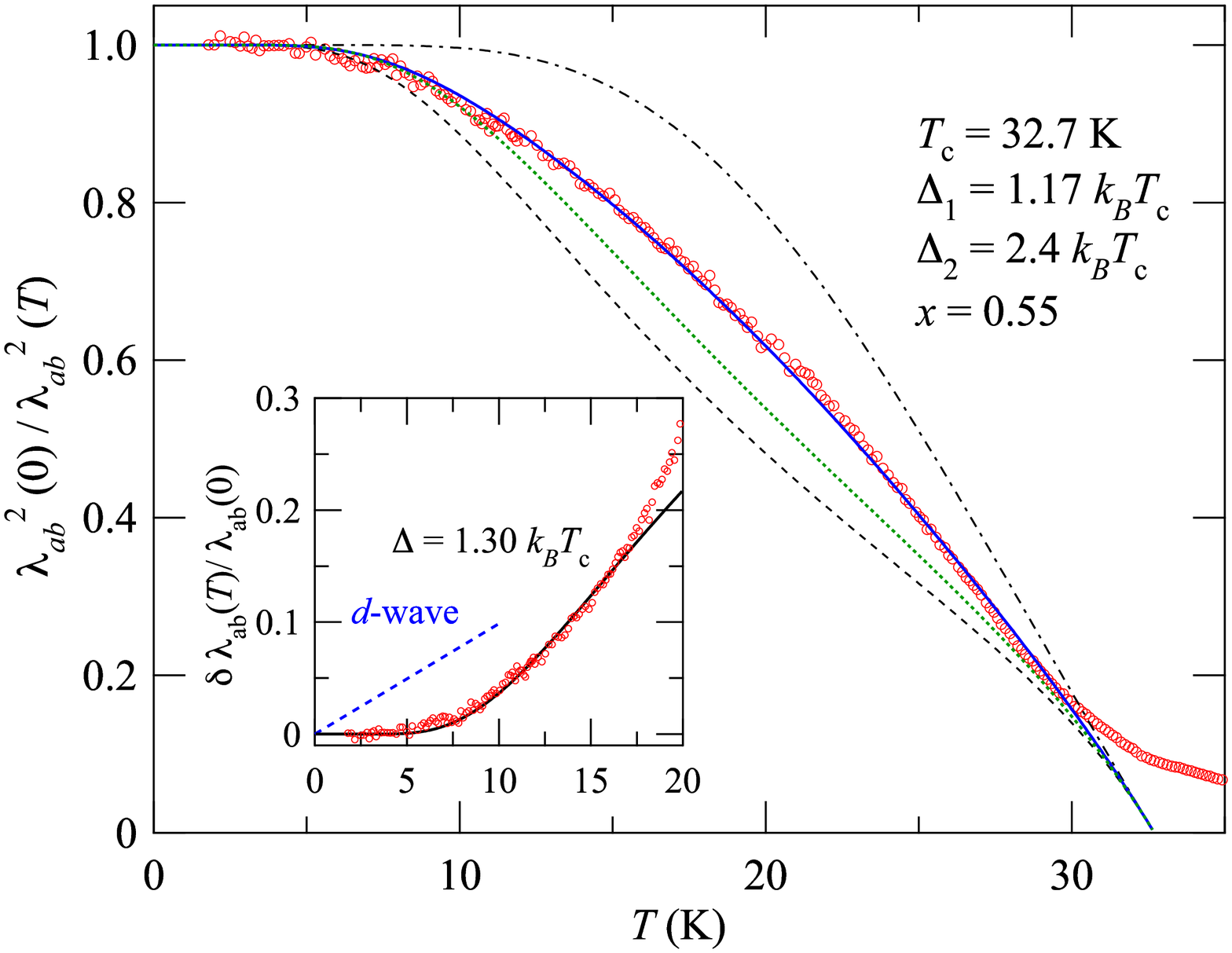}%
\caption{(color online). Inset: $\delta\lambda_{ab}(T)/\lambda_{ab}(0)$ at low temperatures for the cleanest crystal \#3 fitted to Eq.~(\ref{BCS}) with $\Delta=1.30k_BT_c$ (solid line). The dashed line represents $T$-linear dependence expected in clean $d$-wave superconductors \cite{Bon07} with maximum gap $\Delta_0=2k_BT_c$ \cite{Nak08}. Main panel: $\lambda_{ab}^2(0)/\lambda_{ab}^2(T)$ for crystal \#3 fitted to the two-gap model (blue solid line) with $\Delta_1=1.17k_BT_c$ (dashed line) and $\Delta_2=2.40k_BT_c$ (dashed-dotted line). Green dotted line is the single-gap fit using $\Delta=1.30k_BT_c$. Above $T_c$, the normal-state skin depth contribution gives a finite tail.}
\label{lambda}
\end{figure}
%%%%%%%%%%%%%%%%%%%%%%%%%%%%%%%%%%%FIG 3%%%%%%%%%%%%%%%%%%%%

For the cleanest sample, $\delta\lambda_{ab}(T)/\lambda_{ab}(0)$ is depicted in the inset of Fig.~\ref{lambda}, which obviously contradicts the $T$-linear dependence expected in clean $d$-wave superconductors with line nodes. The low-temperature data can rather be fitted to the exponential dependence for full-gap superconductors 
\begin{equation}
\frac{\delta\lambda_{ab}(T)}{\lambda_{ab}(0)} \approx \sqrt{\frac{\pi\Delta}{2k_{\rm B}T}}\exp\left(-\frac{\Delta}{k_{\rm B}T}\right)
\label{BCS}
\end{equation}
with a gap value $\Delta=1.30k_BT_c$. This provides compelling evidence that the intrinsic gap structure in clean samples has no nodes in 122 system. Together with the fact \cite{Has08} that in 1111 $\delta\lambda_{ab}(T)$ also shows exponential behavior, we surmise that both electron and hole-doped Fe-arsenides are intrinsically full-gap superconductors. 

As shown in the main panel of Fig.~\ref{lambda}, the overall temperature dependence of $n_s$ in crystal \#3 cannot be fully reproduced by the single gap calculations. Considering the multiband electronic structure in this system \cite{Zha08,Din08,Evt08}, we fit the data to the two-gap model $n_s(T)=xn_{s1}(T)+(1-x)n_{s2}(T)$ \cite{Bou01}. Here the band 1 (2) has the superfluid density $n_{s1}$ ($n_{s2}$) which is determined by the gap $\Delta_1$ ($\Delta_2$), and $x$ defines the relative weight of each band to $n_{s}$. We obtain an excellent fit with $\Delta_1/k_BT_c=1.17$, $\Delta_{2}/k_BT_c=2.40$ and $x=0.55$. This leads us to conclude the nodeless multi-gap superconductivity having at least two different gaps in this system. It is noteworthy that the obtained gap ratio is comparable to the value $\Delta_2/\Delta_1\approx 2$ found in the ARPES studies \cite{Din08,Evt08} for different bands. A large value of $x=0.55$ implies that the Fermi surface with the smaller gap $\Delta_1$ has a relatively large volume or carrier number, which is also in good correspondence with the ARPES results. 

%%%%%%%%%%%%%%%%%%%%%%%%%%%%%%%%%%%FIG 4%%%%%%%%%%%%%%%%%%%%
\begin{figure}[t]
\includegraphics[width=85mm]{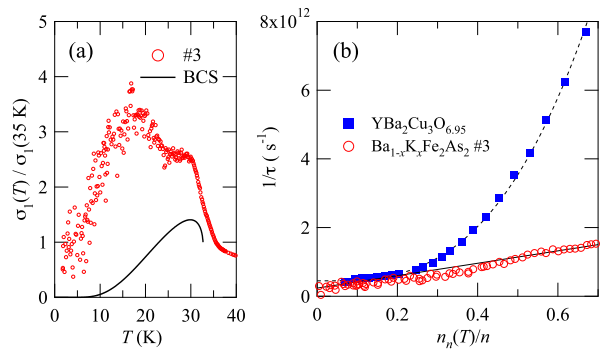}%
\caption{(color online). (a) Temperature dependence of normalized quasiparticle conductivity at 28~GHz for crystal \#3. The solid line is a BCS calculation with $\tau(T_c)=4.4\times10^{-13}$~s. (b) Quasiparticle scattering rate $1/\tau$ as a function of the quasiparticle density $n_n(T)/n$ with a comparison to the results in YBa$_2$Cu$_3$O$_{6.95}$ at 34.8~GHz \cite{Bon94}. The solid and dashed line are fits to the linear and cubic dependence, respectively.} \label{sigma}
\end{figure}
%%%%%%%%%%%%%%%%%%%%%%%%%%%%%%%%%%%FIG 4%%%%%%%%%%%%%%%%%%%%

Next we discuss the low-energy quasiparticle dynamics. In Fig.~\ref{sigma} we show the temperature dependence of the quasiparticle conductivity $\sigma_1(T)/\sigma_1(35$~K) in the cleanest sample \#3. It is evident that below $T_c$, $\sigma_1(T)$ is enhanced from the normal-state values. Near $T_c$, the effects of coherence factors and superconducting fluctuations \cite{Bon07} are known to enhance $\sigma_1(T)$. The former effect, known as a coherence peak, is represented by the solid line in Fig.~\ref{sigma}(a). We note that in the $s_\pm$ pairing state, the coherence peak in the NMR relaxation rate can be suppressed by a partial cancellation of total susceptibility $\sum_{\bf q}{\chi({\bf q})}$ owing to the sign change between the hole and electron bands \cite{Par08,Nag08}. For microwave conductivity, the long wave length limit ${\bf q} \rightarrow 0$ is important and the coherence peak can survive \cite{Dah08}, which may explain the bump in $\sigma_1(T)$ just below $T_c$. At lower temperatures, where the coherence and fluctuation effects should be vanishing, $\sigma_1(T)$ shows a further enhancement. A similar but less pronounced enhancement has been observed in a 1111 system \cite{Has08}. This $\sigma_1(T)$ enhancement can be attributed to the enhanced quasiparticle scattering time $\tau$ below $T_c$. The competition between increasing $\tau$ and decreasing quasiparticle density $n_n(T)=n-n_s(T)$ makes a peak in $\sigma_1(T)$. This behavior is ubiquitous among superconductors having strong inelastic scattering in the normal state \cite{Bon07,Orm02,Shi07}. Following the pioneering work by Bonn {\it et al.} \cite{Bon94}, we employ the two-fluid analysis to extract the quasiparticle scattering rate $1/\tau(T)$ at low temperatures below $\sim 25$~K. In Fig.~\ref{sigma}(b), the extracted $1/\tau(T)$ is plotted against the normalized quasiparticle density $n_n(T)/n$ and compared with the reported results in the $d$-wave superconductor YBa$_2$Cu$_3$O$_{6.95}$ \cite{Bon94}. It is found that the scattering rate scales almost linearly with the quasiparticle density in our 122 system, which is distinct from $1/\tau$ in cuprates that varies more rapidly as $\sim n_n^3$. Such cubic dependence in cuprates is consistent with the $T^3$ dependence of spin-fluctuation inelastic scattering rate expected in $d$-wave superconcuctors, which have $T$-linear dependence of $n_n$ \cite{Qui94}. In $s$-wave superconductors without nodes, $n_n(T)$ and $1/\tau(T)$ are both expected to follow exponential dependence $\sim\exp({-\Delta/k_BT})$ at low temperatures \cite{Qui94}, which leads to the linear relation between $1/\tau(T)$ and $n_n(T)$. So this newly found relation further supports the fully gapped superconductivity in this system. We also note that such analysis yields residual scattering rates at low temperatures for each sample [see Table I], which reinforces the sample-dependent disorder. 

In summary, we measured surface impedance in several Ba$_{1-x}$K$_x$Fe$_2$As$_2$ single crystals. The temperature dependence of the superfluid density depends on the samples having different scattering rates. In the cleanest sample with the smallest $1/\tau$, the superfluid density shows exponential behavior consistent with fully opened two gaps. %This may account for some of the discrepancies found in the literature on the gap structure in the Fe-based superconductors. 
The scattering rate analysis highlights the difference from the $d$-wave cuprates, which also supports the conclusion that the intrinsic order parameter in Fe-As superconductors is nodeless all over the Fermi surface.

%\section*{Acknowledgments}

We thank T. Dahm, H. Ikeda and H. Kontani for discussion. This work was supported by Grant-in-Aid for GCOE program ``The Next Generation of Physics, Spun from Universality and Emergence'' from MEXT, Japan. 

{\it Note added--} After submission, large quasiparticle thermal Hall conductivity in Ba$_{1-x}$K$_x$Fe$_2$As$_2$ has been reported \cite{Ong08}, consistent with the enhanced $\tau(T)$ below $T_c$.


\begin{references}
\bibitem{EDX}
The variation of $x$ measured for different crystals by EDX in $10\times12~\mu$m$^2$ windows is fairly small ($<0.02$), suggesting that the doping level is uniform in this scale.
\end{references}
\end{document}